\documentclass[aps, prd, twocolumn, lengthcheck, superscriptaddress, showpacs, letterpaper, nofootinbib]{revtex4-1}

\usepackage{epsfig}
\usepackage[usenames]{color}
\usepackage{graphicx}
\usepackage{amsmath}
\usepackage{epstopdf}
\usepackage{multirow}
\usepackage{booktabs}
\usepackage{threeparttable}

\newcommand\sect[1]{\emph{#1}---}
\def\bi{\bibitem}

\def\be{\begin{eqnarray}}\def\ee{\end{eqnarray}}
\def\lsim{\mathrel{\rlap{\lower3pt\hbox{\hskip1pt$\sim$}}
     \raise1pt\hbox{$<$}}} 
\def\gsim{\mathrel{\rlap{\lower3pt\hbox{\hskip1pt$\sim$}}
     \raise1pt\hbox{$>$}}} 

\allowdisplaybreaks

\usepackage{multirow}

\allowdisplaybreaks

\begin{document}

\title{Sound velocity and tidal deformability in compact stars}

\author{Yong-Liang Ma}
\email{yongliangma@jlu.edu.cn}
\affiliation{Center for Theoretical Physics and College of Physics, Jilin University, Changchun, 130012, China}

\author{Mannque Rho}
\email{mannque.rho@ipht.fr}
\affiliation{Institut de Physique Th\'eorique, CEA Saclay, 91191 Gif-sur-Yvette c\'edex, France }

\date{\today}

\begin{abstract}
The sound velocity $v_s$ and dimensionless tidal deformability $\Lambda$ are analyzed using the pseudo-conformal model we developed before. In contrast to the conclusion obtained in the previous works in the literature, our model with the upper bound of the sound velocity $v_s = 1/\sqrt{3}$, the so-called conformal sound velocity, set in at a { density relevant to compact stars} $\gsim 2 n_0$ where $n_0$ is the normal nuclear matter density,  can accommodate {\it all} presently established nuclear matter and compact-star properties including the maximum star-mass  constraint  $ \simeq 2.3 M_\odot$. This observation is associated with a possible emergence of pseudoconformal structure in compact star matter---in which the trace of energy-momentum tensor is a nearly density-independent nonzero constant---brought in by a topology change at  $2.0 \lesssim n_{1/2}/n_0 \lesssim 4.0$ commensurate with a possible change of degrees of freedom from hadrons.
\end{abstract}

\maketitle

\section{The Problem}

The equation of state (EoS) of dense nuclear matter is an extremely important but not  well-understood subject in nuclear physics and astrophysics. It is closely related to the phase structure of QCD under extreme conditions. At this moment, it is well accepted that nuclear matter up to the nuclear saturation density $n_0$ can be described by using the equation of state derived from such hadrodynamics as effective density functionals, chiral effective field approaches and others. However, when the density of nuclear matter is increased to $\gtrsim 2 n_0$, the situation is totally unclear. It is exciting that the possible empirical constraint has begun to come from the observation of compact stars, especially in the present post-Newton era since the observation of a binary neutron star merger~\cite{GW170817}.

Recently, an issue concerning the upper bound of the sound velocity $v_s$ was raised in the theoretical construction of the EoS at high baryon density. Naive causality consideration sets the bound to $v_s \leq 1$ (in units of $c=1$). However, if we consider an extreme case in which the nuclear matter is made of ultrarelativistic massless particles, the constraint is brought down to $v_s \leq 1/\sqrt{3}$,  the upper bound in a system with conformal symmetry. Whether or not such an upper bound can be saturated in dense nuclear matter should be settled by observation. A recent analysis in Ref.~\cite{Bedaque:2014sqa} argues that, in combination with the EoS of hadronic matter that describes correctly the nuclear matter properties around $n_0$, the existence of neutron stars with mass $\sim 2.0 M_\odot$~\cite{Nat-2solar,Sci-2solar} is not consistent with bound $v_s \leq 1/\sqrt{3}$. This ``no-go" argument has been given support by several other works, for example， Refs.~\cite{tews,moustakidis,Alsing:2017bbc}.  Furthermore exceeding  from the conformal velocity in certain standard nuclear model results has been suggested to be  associated with a generic feature of scalar extensions of general relativity ~\cite{TEMT}.

The purpose of this note is to show that, contrary to the no-go arguments cited above which seems to be well accepted by the general community, the pseudoconformal model (PCM) we proposed in Refs.~\cite{PKLMR,MLPR,MR-PCM}---which is drastically different---is fully consistent with the present observation of massive neutron stars and tidal deformability in the detection of gravitation waves~\cite{abottetal}. Whether or not this scenario will survive the  scrutiny from forthcoming astrophysical data is an open issue for the future.

The PCM for dense nuclear matter we proposed is based on the scale-chiral effective theory of QCD in which, in addition to the pseudo-Numbu-Goldstone boson pion, the nucleon field figures as the matter field as in the standard chiral perturbation theory, the lowest-lying vector mesons $\rho$ and $\omega$ are incorporated as hidden local symmetry (HLS) fields and a light scalar meson is introduced as the dilaton. We call this $bs$HLS Lagrangian. Once this $bs$HLS is matched to QCD via correlators~\cite{HY:WilsonMatch}, its low energy constants can be endowed with the condensates of QCD such as $\langle\bar{q}q\rangle$, $\langle G^2\rangle$, etc. The condensates of QCD should get modified when the vacuum is changed by the density, rendering the parameters of the effective theory density dependent.  This density dependence, referred to as ``intrinsic density dependence (IDD)," gets additional density-dependent corrections from  renormalization group (RG) decimation from the matching scale down to an effective nuclear interaction scale in baryonic matter, referred to as   ``induced density dependence (DD$_{\rm induced}$)."  The combined density dependence will be denoted as IDD$^\ast$. The resulting effective field theory ~\cite{PKLMR,MLPR,MR-PCM} encodes the following elements in the effective density functional in modeling QCD that the approach provides.
\begin{itemize}
  \item Topology: Topology figures in the description of nuclear matter at high density in which the nucleons are compressed such that the nuclear matter can be regarded as a crystal matter. In QCD, baryons at high density can be described as skyrmion matter put on crystal lattice. This is justified in the large $N_c$ limit (for reviews, see e.g., Refs.~\cite{park-vento,MR-review}). A robust conclusion obtained in the skyrmion approach, absent in nontopological approaches, i.e., standard chiral perturbation theory,  is the topology change from the crystal matter  of skyrmions into a  crystal matter of half-skyrmions. The density at which this topology change takes place is denoted as  $n_{1/2}$.  This density, thus far accessible neither from theory nor from experiments, is found to be in the range $2\lsim n_{1/2}/n_0\lsim 4$. This comes from the analysis of the sound velocity of stars in conjunction with the tidal deformability~\cite{MLPR,MR-PCM}.

The most crucial outcome of the present work is that in going from the skyrmion matter to the half-skyrmion matter, parity doubling in the nucleon structure arises~\cite{maetal-doublet}. Parity doubling is not in QCD in the matter-free vacuum, so the process involved here is an emergence of hidden symmetry of QCD. It exposes the possible origin of the nucleon mass. The nucleon mass tends to a constant $m_0\sim (0.6-0.9)m_N$ where $m_N$ is the free-space nucleon mass and remains density-independent constant up to a putative deconfinement density. This observation has an important consequence on the trace of energy-momentum tensor as mentioned below.

\item Scale symmetry: The scale symmetry provides a powerful access to the scalar meson with a mass $\sim 600~$MeV in an effective field theory~\cite{CT}. It can be regarded as the Nambu-Goldstone boson, dilaton $\chi$, of the spontaneous breaking of scale symmetry triggered by its explicit breaking, i.e., the trace anomaly. When the trace anomaly is totally attributed to the dilaton potential term, which appears to be a fairly good approximation in low-energy nuclear dynamics~\cite{PKLMR,gA}, one can write
      \begin{eqnarray}
      V(\chi) & = & \frac{m_\chi^2 f_\chi^2}{4}\left(\frac{\chi}{f_\chi}\right)^4 \left(\ln\frac{\chi}{f_\chi} - \frac{1}{4} \right)
      \end{eqnarray}
      which yields
      \begin{eqnarray}
      \langle \delta V \rangle & = & {} - \langle \theta_\mu^\mu \rangle = \frac{m_\chi^2}{4 f_\chi^2} \langle \chi^4 \rangle.
      \end{eqnarray}

  \item Local flavor symmetry: The local flavor symmetry provides a powerful approach to include the vector mesons in the chiral effective field theory through the gauge principle such that both the vector mesons and the pion can be treated on the same footing.  Although this local flavor symmetry is broken in the matter-free vacuum, renormalization-group analysis shows that at high density   the $\rho$-nucleon coupling goes to zero, together with the $m_\rho$ going to zero~\cite{HY:VM1,HY:VM2}.

\end{itemize}

Implementing the above principal features into the EFT Lagrangian $bs$HLS, the topology change  density $n_{1/2}$ provides the separation between Region-I (R-I) for the density regime $n < n_{1/2}$ and Region-II (R-II) for $n \geq n_{1/2}$.  In R-I, the $bs$HLS is essentially equivalent to the standard EFT (SEFT) anchored on chiral symmetry---with the advantage of having the vector mesons appearing at the leading order and the IDD$^\ast$ in the parameters, which in SEFT would require going to higher orders. In R-II, however, with the input from the topology change, the structure is drastically different from  the SEFT with the effects from emerging symmetries. What is also important in R-II is that a possible change of degrees of freedom is encoded therein. As argued in \cite{PKLMR,MR-PCM}, this region could mimic the changeover from baryons to strongly-coupled quarks and gluons as described in \cite{baymetal}. This connection could be  considered as a Cheshire-Cat phenomena developed in 1980s~\cite{MR-PCM,cheshirecat}

\section{The pseudo-conformal model (PCM)}
As a consequence of the parity-doubling together with the  intervention of the hidden symmetries (the scale and flavor local symmetries),  the vacuum expectation value of the trace of the energy momentum tensor (TEMT) $\theta_\mu^\mu$ becomes a nearly\footnote{We say nearly because we are ignoring the light-quark masses. We have not verified that there is no density dependence arising from the chiral symmetry-breaking effect. We are reasonably certain, however,  that such a density dependence, if there is any, would not be important since $m_0$ is quite substantially big compared with the pion mass.}  density-independent nonzero constant in R-II, i.e., $n \gtrsim n_{1/2}
$\begin{eqnarray}
\langle\theta_\mu^\mu \rangle & = & \epsilon - 3 P \propto f(m_0)
\end{eqnarray}
with $m_0$ being the chiral invariant nucleon mass. Therefore, the derivative with respect to density of  $\theta_\mu^\mu$
\begin{eqnarray}
\frac{\partial}{\partial n}\langle\theta_\mu^\mu \rangle & = & \frac{\partial \epsilon (n)}{\partial n} (1-3{v_s^2})\label{derivTEMT}
\end{eqnarray}
(where  $v_s^2=\frac{\partial P(n)}{\partial n}/\frac{\partial \epsilon(n)}{\partial n}$) vanishes.  Since $\frac{\partial \epsilon (n)}{\partial n} \neq 0$ in the range of densities involved, we immediately obtain
\begin{eqnarray}
v_s = \frac{1}{\sqrt{3}},
\end{eqnarray}
which is usually associated with the sound velocity in a conformal symmetric system. Here, however, since the TEMT is not equal to zero, we call it  ``pseudoconformal sound velocity."

It was shown in \cite{PKLMR} that the pseudoconformal sound velocity comes out in the highly successful $V_{lowk}$ renormalization group approach~\cite{VlowK} based on $bs$HLS described above for $n_{1/2}=2.0n_0$. The full $V_{lowk}$ RG result is exactly reproduced by R-I described by the $V_{lowk}$RG treatment and by R-II given by the formula of pseudoconformality
\begin{eqnarray}
E/A  = - m_N +X^\alpha  x^{1/3} + Y^\alpha x^{-1}, ~ & \hbox{$n \gtrsim n_{1/2}$.}
\label{eq:EoSPCM}
\end{eqnarray}
where $x = n/n_0$ and $X, Y$  are parameters given by the pressure and chemical potential matched between R-I (given by $V_{lowk}$) and R-II (given by (\ref{eq:EoSPCM}))  at $n_{1/2}$  for  $\alpha=(N-Z)/(N+Z)$. 
We assume that this PCM is applicable for $n_{1/2}> 2n_0$.

\begin{figure}[h]
 \begin{center}
   \includegraphics[width=7cm]{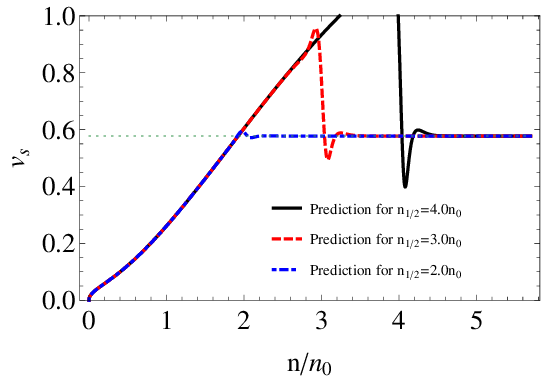}
   \includegraphics[width=7cm]{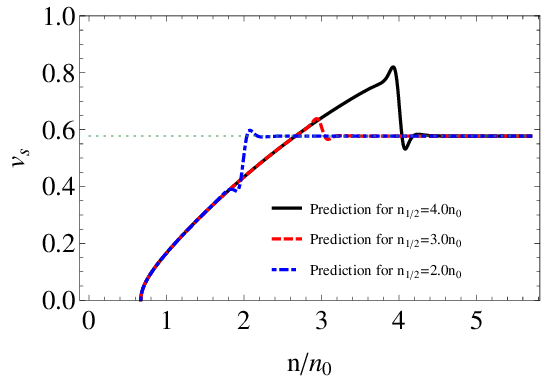}
  \end{center}
 \caption{Sound velocity as a function of density in neutron matter (upper panel) and symmetric matter (lower panel).}
\label{fig:vs}
\end{figure}

\section{Star properties}
We first look at the density dependence of the sound velocity. Our results are plotted in Fig.~\ref{fig:vs}. From this figure, we see that at high density, in all three cases of $n_{1/2}$, $v_s \to 1/\sqrt{3}$, the conformal limit. But the sound velocities  are different at intermediate densities for different values  of $n_{1/2}$. Especially, for $n_{1/2} \simeq 4.0 n_0$, $v_s > 1$, violating causality. This reinforces the upper bound  $n_{1/2} < 4.0 n_0$ arrived at with the pressure bound given by presently available heavy-ion data in Ref.~\cite{MR-PCM}.

Figure~\ref{fig:vs} tells us that for $n < 2.0 n_0$, the three choices of $n_{1/2}$ yield the same sound velocity. Therefore, to see the correlation between the sound velocity and the tidal deformability, the central density of the neutron star should be $\gtrsim 2.0 n_0$. Explicitly, to see the correlation between the sound velocities for $n_{1/2} = 2.0 n_0$ and $3.0 n_0$ and the tidal deformability, the central density of the neutron star should be $\gtrsim 2.0 n_0$ while to see the correlation between the sound velocities for $n_{1/2} = 3.0 n_0$ and $4.0 n_0$ and the tidal deformability, the central density of the neutron star should be $\gtrsim 3.0 n_0$.

Now we turn to the tidal deformability in conjunction with the sound velocity. Listed in
Table.~\ref{tab:StarProperty} are the star properties for different massive stars. We see that, for the star $M_{1.4}$ (here and in what follows  the subindex stands for the mass of the star {in unit of solar mass}), since the central density is $2.0n_0 < n_{cent} \leq 2.3 n_0$, when $n_{1/2}$ is changed from $2.0n_0$ to $3.0 n_0$, the tidal deformability is reduced and the size  shrinks. However, when $n_{1/2}$ is changed from $3.0n_0$ to $4.0 n_0$, neither the tidal deformability nor the star size is affected. A similar analysis applies to other stars in the table. It is surprising that the tidal deformability and the radius of the star are nearly independent of the location of $n_{1/2} > 2n_0$.

\begin{widetext}

\begin{table}[!ht]
\centering
\begin{threeparttable}
\caption{Properties of compact stars with different masses and $n_{1/2}/n_0$.}
\label{tab:StarProperty}
\begin{tabular}{cccccccccc}
\hline
\hline
\multirow{2}{*}{$M/M_\odot$}& \multicolumn{3}{c}{$n_{cent}/n_0$}&\multicolumn{3}{c}{$\Lambda/100$}&\multicolumn{3}{c}{$R$/km}\cr
\cmidrule(lr){2-4}\cmidrule(lr){5-7}\cmidrule(lr){8-10}
&$n_{1/2}=2.0$&$n_{1/2}=3.0$&$n_{1/2}=4.0$&$n_{1/2}=2.0$&$n_{1/2}=3.0$&$n_{1/2}=4.0$&$n_{1/2}=2.0$&$n_{1/2}=3.0$&$n_{1/2}=4.0$
\cr
\hline
1.40 &2.02&2.30&2.30&7.85&6.52&6.52&13.0&12.8&12.8\cr\hline
1.60 &2.61&2.54&2.54&2.85&2.90&2.90&12.8&12.8&12.8\cr\hline
1.80 &3.11&2.84&2.81&1.21&1.30&1.30&12.8&12.8&12.8\cr\hline
2.00 &4.50&3.60&3.21&0.37&0.55&0.55&11.5/12.2&12.6&12.7\cr\hline
2.20 & $\cdots$ & $\cdots$ &4.00& $\cdots$ & $\cdots$ &0.20& $\cdots$ & $\cdots$ &12.3\cr
\hline
\hline
\end{tabular}
\end{threeparttable}
\end{table}
\end{widetext}

\section{Discussion}
We have found, as summarized in this note, that in contrast to what was found in the literature, the PCM,  which works well for normal nuclear matter density, gives $v_s \to 1/\sqrt{3}$---the conformal limit---at a density $\gsim n_{1/2} $ and accommodates massive neutron stars up to $2.23 M_\odot$, which is consistent with the present observation~\footnote{There are approaches in the literature that hybridize hadronic models at low density and quark models at high density~\cite{Alford:2013aca}. Such models inevitably include phase transitions and consequently different predictions.}. In addition to the constraint from the massive star, the tidal deformability obtained is consistent with the  presently available empirical bound from the gravitational wave observation both for $\Lambda_{1.4}$ and radius $R_{1.4}$. The information coming from the sound velocity and tidal deformability pins down the topology change density to $2.0 n_0 < n_{1/2} < 4.0 n_0$. This can be taken as a signal for the change of degrees of freedom from normal baryon (skyrmion) to an ``exotic" fermion (half-skyrmion), such a change of degrees of freedom being required to accommodate the (pseudo-)conformal velocity as argued in \cite{tews}.  The topology change could be a dual picture to the baryons-to-quark continuity argued to be present at the same range of density.

\begin{figure}[h]
 \begin{center}
   \includegraphics[width=8cm]{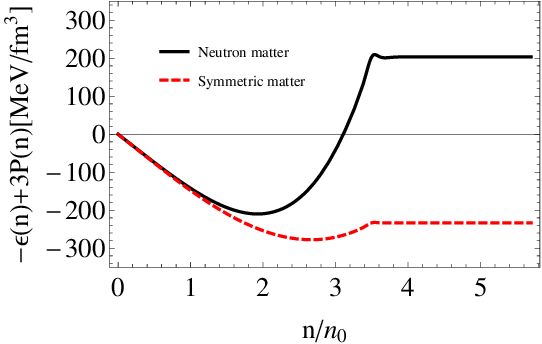}
  \end{center}
 \caption{Trace of the energy momentum tensor as a function of density with $n_{1/2} = 3.5 n_0$ in PCM.}
\label{fig:TEMT35}
\end{figure}

Although thus far our pseudoconformal structure at $n>n_{1/2}\sim 2n_0$ is not torpedoed by theory or experiments, it is so drastically different from the standard one typified by chiral effective field theories with baryons and (pseudo-)Goldstone bosons as the only relevant degrees of freedom and also various sophisticated energy density functionals that we are compelled to search for what could obstruct the PCM scenario. One near future source for such an obstruction could be the on-going LIGO/Virgo observations of gravity waves.
Should the $\Lambda_{1.4}$ be tightened to a much smaller value than what we have, i.e., $650$, the PCM giving the pseudoconformal velocity, would be in tension. At the moment going down a lot more seems difficult to accommodate. Another is the observation made by Ref.~\cite{TEMT} where it has been argued that there will be a sign change from negative to  positive of the TEMT in scalar-tensor theories of gravity at high density relevant to massive compact stars and the positiveness is related to the condition that $v_s^2 > 1/3$. This observation would be vindicated by the analyses of~\cite{Bedaque:2014sqa,tews,moustakidis,Alsing:2017bbc} which disfavor the bound $v_s^2 \leq 1/3$. Our model gives a counter-example to this possibility. In our PCM model, when $n_{1/2}$ is larger than certain value, say,  $n_{1/2} \gtrsim 3.0 n_0$, there is a sign change in the TEMP,  becoming positive at $n_{1/2}$. This is  shown in Fig.~\ref{fig:TEMT35} for $n_{1/2} =3.5n_0$. {In addition,  as argued in Ref.~\cite{tews},  the conformal speed of sound could be approached {\it only} at an asymptotic density $> 50 n_0$ due to the vanishing of the TEMT approaching the UV fixed point of QCD. However, in our PCM, the density independence of the TEMT---in the range of densities relevant to compact stars---makes the sound speed pseudoconformal for $n>n_{1/2}$. How our PC sound speed at $n\sim (3-7)n_0$ goes over to the truly conformal sound speed at $n > 50n_0$ (approaching perhaps the color-flavor-locked state) remains to be understood.

We finally want to say is that, although it is straightforward to extend our framework to include strangeness which has been discussed for several decades, we will not cover it in the work because, due to the paucity of experimental information and consistent theoretical tools up to date, there is no generally accepted scenarios (see e.g., Ref.~\cite{CNDIII} for a comprehensive discussion.)

\acknowledgments

We grateful to Hyun Kyu Lee for pointing out Ref.~\cite{TEMT} for the sign change in the TEMT  and for helpful discussions on its relevance to the pseudo-conformal model.
Y.~L. Ma was supported in part by National Science Foundation of China (NSFC) under Grants No.~ 11875147, No.~11475071, No.~11747308 and the Seeds Funding of Jilin University. 

\sect{Note added}
After this paper was submitted, Jeong, Mclerran and Sen suggest to regard the Quarkyonic matter as a Fermi sphere of quark matter surrounded by a shell of nucleon in matter~\cite{Jeong:2019lhv}.  Both pictures agree that there are deconfined quasiparticles at $(2-4) n_0$, ours in terms of half-skyrmions, most likely  deconfined, and Ref.~\cite{Jeong:2019lhv} in terms of deconfied quarks as constituent quarks.
In addition, in both pictures, the sound velocity could approach to $v_s^2 \to 1/3$---the conformal limit---at $n > 2.0 n_0$.

\end{document}